\documentclass[aps,prl,a4paper,superscriptaddress,amsmath,amssymb,twocolumn,showpacs]{revtex4} 
\usepackage{graphicx,epsfig}
\usepackage[usenames]{color}

\allowdisplaybreaks
\begin{document}

\newcommand{\Avg}[1]{\langle \,#1\,\rangle_G}

\newcommand{\E}{\mathcal{E}}
\newcommand{\Lag}{\mathcal{L}}
\newcommand{\M}{\mathcal{M}}
\newcommand{\N}{\mathcal{N}}
\newcommand{\U}{\mathcal{U}}
\newcommand{\R}{\mathcal{R}}
\newcommand{\F}{\mathcal{F}}
\newcommand{\V}{\mathcal{V}}
\newcommand{\C}{\mathcal{C}}
\newcommand{\I}{\mathcal{I}}
\newcommand{\s}{\sigma}
\newcommand{\up}{\uparrow}
\newcommand{\dw}{\downarrow}
\newcommand{\h}{\mathcal{H}}
\newcommand{\Hn}{\mathcal{H}}
\newcommand{\himp}{\hat{h}}
\newcommand{\g}{\mathcal{G}^{-1}_0}
\newcommand{\D}{\mathcal{D}}
\newcommand{\A}{\mathcal{A}}
\newcommand{\projs}{\hat{\mathcal{S}}_d}
\newcommand{\proj}{\hat{\mathcal{P}}_d}
\newcommand{\K}{\textbf{k}}
\newcommand{\Q}{\textbf{q}}
\newcommand{\hzero}{\hat{T}}
\newcommand{\io}{i\omega_n}
\newcommand{\eps}{\varepsilon}
\newcommand{\+}{\dag}
\newcommand{\su}{\uparrow}
\newcommand{\giu}{\downarrow}
\newcommand{\0}[1]{\textbf{#1}}
\newcommand{\ca}{c^{\phantom{\dagger}}}
\newcommand{\cc}{c^\dagger}
\newcommand{\aaa}{a^{\phantom{\dagger}}}
\newcommand{\aac}{a^\dagger}
\newcommand{\bba}{b^{\phantom{\dagger}}}
\newcommand{\bbc}{b^\dagger}
\newcommand{\da}{d^{\phantom{\dagger}}}
\newcommand{\dc}{d^\dagger}
\newcommand{\fa}{f^{\phantom{\dagger}}}
\newcommand{\fc}{f^\dagger}
\newcommand{\ha}{h^{\phantom{\dagger}}}
\newcommand{\hc}{h^\dagger}
\newcommand{\be}{\begin{equation}}
\newcommand{\ee}{\end{equation}}
\newcommand{\bea}{\begin{eqnarray}}
\newcommand{\eea}{\end{eqnarray}}
\newcommand{\ba}{\begin{eqnarray*}}
\newcommand{\ea}{\end{eqnarray*}}
\newcommand{\dagga}{{\phantom{\dagger}}}
\newcommand{\bR}{\mathbf{R}}
\newcommand{\bQ}{\mathbf{Q}}
\newcommand{\bq}{\mathbf{q}}
\newcommand{\bqp}{\mathbf{q'}}
\newcommand{\bk}{\mathbf{k}}
\newcommand{\bh}{\mathbf{h}}
\newcommand{\bkp}{\mathbf{k'}}
\newcommand{\bp}{\mathbf{p}}
\newcommand{\bL}{\mathbf{L}}
\newcommand{\bRp}{\mathbf{R'}}
\newcommand{\bx}{\mathbf{x}}
\newcommand{\by}{\mathbf{y}}
\newcommand{\bz}{\mathbf{z}}
\newcommand{\br}{\mathbf{r}}
\newcommand{\Ima}{{\Im m}}
\newcommand{\Rea}{{\Re e}}
\newcommand{\Pj}[2]{|#1\rangle\langle #2|}
\newcommand{\ket}[1]{\vert#1\rangle}
\newcommand{\bra}[1]{\langle#1\vert}
\newcommand{\setof}[1]{\left\{#1\right\}}
\newcommand{\fract}[2]{\frac{\displaystyle #1}{\displaystyle #2}}
\newcommand{\Av}[2]{\langle #1|\,#2\,|#1\rangle}
\newcommand{\av}[1]{\langle #1 \rangle}
\newcommand{\Mel}[3]{\langle #1|#2\,|#3\rangle}
\newcommand{\Avs}[1]{\langle \,#1\,\rangle_0}
\newcommand{\eqn}[1]{(\ref{#1})}
\newcommand{\Tr}{\mathrm{Tr}}

\newcommand{\Pg}{\mathcal{P}_G}

\newcommand{\Vb}{\bar{\mathcal{V}}}
\newcommand{\Vd}{\Delta\mathcal{V}}
\newcommand{\Pb}{\bar{P}_{02}}
\newcommand{\Pd}{\Delta P_{02}}
\newcommand{\tb}{\bar{\theta}_{02}}
\newcommand{\td}{\Delta \theta_{02}}
\newcommand{\Rb}{\bar{R}}
\newcommand{\Rd}{\Delta R}

\title{Interplay of spin-orbit and entropic effects in Cerium}

\author{Nicola Lanat\`a}
\affiliation{Department of Physics and Astronomy, Rutgers University, Piscataway, New Jersey 08856-8019, USA}
\author{Yong-Xin Yao}
\affiliation{Ames Laboratory-U.S. DOE and Department of Physics and Astronomy, Iowa State
University, Ames, Iowa IA 50011, USA}
\author{Cai-Zhuang Wang}
\affiliation{Ames Laboratory-U.S. DOE and Department of Physics and Astronomy, Iowa State
University, Ames, Iowa IA 50011, USA}
\author{Kai-Ming~Ho}
\affiliation{Ames Laboratory-U.S. DOE and Department of Physics and Astronomy, Iowa State
University, Ames, Iowa IA 50011, USA}
\author{Gabriel Kotliar}
\affiliation{Department of Physics and Astronomy, Rutgers University, Piscataway, New Jersey 08856-8019, USA}

\date{\today} 

\begin{abstract}

We perform
first-principle calculations of elemental cerium
and compute its pressure-temperature phase diagram,
finding good quantitative agreement with the experiments.
Our calculations indicate that,
while a signature of the volume-collapse transition appears
in the free energy already at low temperatures,
at larger temperatures this signature is enhanced because
of the entropic effects, and originates an actual thermodynamical
instability.
Furthermore, we find that the catalyst 
determining this feature is --- in all temperature regimes ---
a pressure-induced effective
reduction of the $f$-level degeneracy due to the spin-orbit coupling.
Our analysis suggests also that
the lattice vibrations might be crucial in order to capture
the behavior of the pressure-temperature transition line
at large temperatures.

\end{abstract}

\maketitle

At ambient temperature elemental cerium undergoes a
pressure-induced first-order isostructural transition,
which is accompanied by a volume collapse of about $14\%$.
The critical pressure where this transition occurs
increases linearly with the temperature, and
the corresponding transition line in the pressure-temperature
phase diagram ends at a critical temperature $T_{\text{c}}\simeq 500\,K$.
Since 1949, when the $\gamma$-$\alpha$ transition was discovered,
it has stimulated a lot of experimental and theoretical work,
but a complete theoretical description of this phenomenon
is still lacking.
In particular,
two theoretical pictures are still nowadays under debate:
(i) the Kondo volume collapse model (KVC), that
was proposed by Allen and Martin~\cite{Allen82} and
independently by Lavagna, Lacroix,
and Cyrot~\cite{Lavagna82}, and 
(ii) the orbital-selective Mott transition (HM),
that was proposed by Johansson~\cite{Johansson74}.
The main difference between these two models is that,
while the KVC attributes the $\gamma$-$\alpha$ transition of
cerium to a rapid change of the degree of hybridization
between the $f$ electrons and the $spd$
conduction bands (and consequently of the Kondo temperature),
within the HM transition the key quantity whose rapid variation
is repsonsible of the transition is the kinetic energy of the 
$f$ electrons, while the itinerant electrons are inert spectators
of the transition.
Consistently with both the HM and the KVC,
photoemission experiments~\cite{Wieliczka,Weschke,Patthey}
have demonstrated that the $f$ electrons are correlated both
in the $\gamma$ phase and in the $\alpha$ phase.
On the other hand, 
calculations of the optical spectrum~\cite{haule-Ce-2005}
have enabled to interpret
the experiments~\cite{Experiments-optical-spectroscopy_Ce}
in favor of the KVC.

In our view, there are few fundamental questions
that have to be answered in order
fully understand the physics
underlying the $\gamma$-$\alpha$ transition or cerium.
(1) Why does $T_K$ vary rapidly only within a
narrow window of volumes rather than varying smoothly?
(2) How does the rapid variation of $T_K$ affect the free energy
and, in particular,
generate the $\gamma$-$\alpha$ isostructural transition?
(3) Is the fact that the $\gamma$-$\alpha$ isostructural transition
ends at $T_{\text{c}}\simeq 500\,K$ purely due to electronic effects
or does it require to take into account the temperature-induced
lattice vibrations?

Recent first principle calculations of cerium~\cite{Our-Ce}
within the Gutzwiller approximation in combination with
the local density approximation (LDA+GA) have suggested
a possible solution to the first of the questions
listed above.
In fact,
these calculations have shown
that a clear signature of the $\gamma$-$\alpha$
isostructural transition of cerium can be observed
at zero temperature if (and only if) the spin-orbit coupling
is taken into account.
Furthermore, they have shown that
the reason underlying this result is that
the spin-orbit coupling effectively reduces the
$f$-level degeneracy from $14$ to $6$ when the hybridization
strength is reduced by increasing the volume, thus causing
a crossover in the evolution of $T_K$.
On the contrary, if the spin-orbit couping is not taken into account
$T_K$ evolves smoothly as a function of the volume.
This point of view has been further clarified in
a following work~\cite{pmee}
making use of the ``principle of maximum entanglement entropy''.
The importance of the spin-orbit coupling for the determination
of the thermodynamical properties of cerium has been finally confirmed
also within the framework
of LDA in combination with
dynamical mean field theory (LDA+DMFT)~\cite{LDA+U+DMFT},
see Ref.~\cite{DMFT-Ce-Amadon},
where the spin-orbit coupling was neglected
because of the great computational complexity of the problem.
In fact, the LDA+DMFT low-temperature pressure-volume phase diagram
--- that is in very good quantitative agreement
with the LDA+GA calculations of Ref.~\cite{Our-Ce} ---
does not display any signature of the transition at low temperatures.
Furthermore, it was shown that,
if the spin-orbit coupling is neglected, 
not even the entropy is sufficient
to induce a thermodynamical instability,
but only a softening of the bulk modulus.

The second of the questions listed above is particularly challenging, as
a rapid variation of $T_K$ affects differently the internal energy $E$ 
and the entropy $S$ of the system.
In fact, at small volumes (where $T_K \gg T$) the Kondo-stabilization
effect decreases the internal-energy~\cite{McMahan,Georges06},
while the entropy is small; on the contrary,
at large volumes (where $T_K \ll T$) the entropy of the system is
large, thus decreasing considerably the free energy and stabilizing
the $\gamma$ phase.
Because of these reasons, it is rather complicated to predict
quantitatively the behavior of the free energy $F=E-TS$
in concomitance with a rapid variation of $T_K$, and accurate 
first-principles calculations are essential in order 
to study this problem.

Since the spin-orbit coupling and
the entropic effects are both important at finite temperatures,
it is clear 
that in order to fully understand the $\gamma$-$\alpha$ transition at
finite temperatures (and in particular to
answer the second and the third questions listed above)
it is essential to perform first principles calculations
able to take into account both of these physical effects.
In this work,
besides taking into account the spin-orbit coupling,
we describe the entropic effects at finite temperatures by
employing a charge self-consistent combination of LDA 
with the rotationally-invariant slave-boson (SB)
mean-field theory~\cite{Lechermann_rotationally-invariant-SB},
that we have implemented following Ref.~\cite{DMFT-GA}.
Note that this method is equivalent to LDA+GA at zero temperature~\cite{equivalence-SB-GA}
and, although it is not as accurate as LDA+DMFT,
it has the advantage to be much less computationally demanding.
For completeness, in the supplemental material
is given a brief introduction of the LDA+SB method.

As in Ref.~\cite{Our-Ce},
we employ the ``standard'' prescription for
the double-counting functional 
and the general (rotationally-invariant)
Slater-Condon parametrization~\cite{LDA+U} of
the on-site interaction,
assuming that the Hund's coupling constant is $J=0.7\,eV$ and that
the value of the interaction-strength is $U=6\,eV$.

\begin{figure}
\begin{center}
\includegraphics[width=8.6cm]{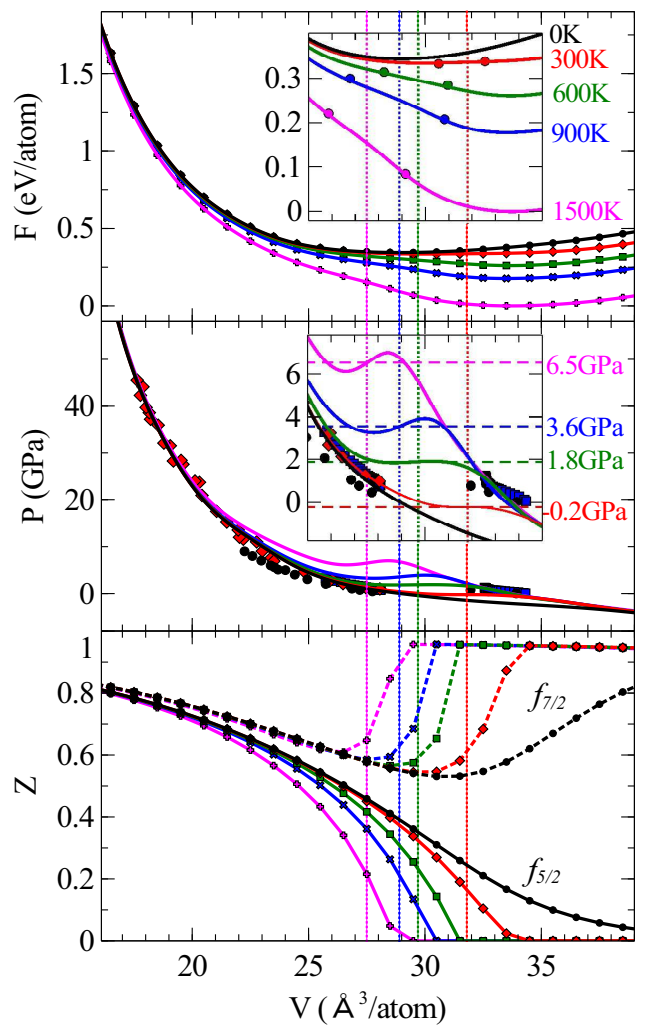}
\caption{Evolution of the free energy (upper panel), pressure (middle panel)
  and Gutzwiller quasi-particle weights (lower panel) as a function of the volume
  at different temperatures. The vertical lines indicate the crossover
  volumes $V_{\text{c}}$ where the pressure $dP/dV$ achieves it's maximum
  (positive) value.
  The dots in the inset of the upper panel indicate the boundaries of the
  volume-collapse transition according to the common-tangent construction.
}
\label{figure1}
\end{center}
\end{figure}
In the upper panel of Fig.~\ref{figure1} is shown the evolution
of the free energy $F=E-TS$ at different temperatures $T$.
While at small volumes
$F$ is not very sensitive to $T$,
at larger volumes it is considerably reduced by the temperature.
Because of this effect,
at $T\gtrsim 300\,K$ the $\gamma$ phase is stable at zero pressure,
while at lower temperatures
the $\gamma$ phase is stable only at negative pressures.
In other words, consistently with the experiments,
our calculations indicate that at $T\gtrsim 300\,K$
the volume-collapse transition occurs at positive pressures.
The boundaries of the transition
have been evaluated within the common-tangent construction,
and are indicated by the dots in the inset of
the upper panel of Fig.~\ref{figure1}.

The pressure-volume curves,
that are calculated from the free energy $F$ using $P=-dF/dV$,
are shown in the middle panel of Fig.~\ref{figure1}
in comparison with the room-temperature experimental data
of Refs.~\cite{CePV77,CePV85}.
The critical pressures (at different temperatures) are here
identified by the equal-area construction,
and are indicated by horizontal lines.
The volumes $V_{\text{c}}$ where $dP/dV$ reaches the maximum value
are indicated by vertical lines.
Remarkably, 
our finite-temperature calculations 
are in very good agreement with the
room-temperature experimental pressure-volume phase diagram.
In particular, note that the entropic effects improve the
comparison with the experiments at large volumes,
which confirms that they are
very important to describe the thermodynamical properties of
cerium, especially in the $\gamma$ phase.

In the lower panel of Fig.~\ref{figure1}
is shown the evolution as a function of the volume
of the Gutzwiller quasi-particle renormalization weights of the
$7/2$ and $5/2$ $f$-electrons.
Interestingly, 
the overall qualitative behavior of the quasi-particle weights is
similar for all temperatures:
both $Z_{5/2}$ and $Z_{7/2}$ are significantly smaller than $1$
already at high pressures,
and in this regime they monotonically decrease as a function of
the volume; on the contrary,
at volumes larger than a temperature-dependent crossover
volume $V_{\text{c}}$ we observe that $Z_{7/2}$ increases
while $Z_{5/2}$ rapidly decreases.
This behavior was already discussed in the zero-temperature
calculations of Ref.~\cite{Our-Ce}, and
reflects the above-mentioned effective reduction of the
$f$-level degeneracy from $14$ to $6$ due to
the spin-orbit coupling.
Remarkably, here we find that
the crossover of the $Z$'s takes place for all temperatures
at the same volume $V_{\text{c}}$ where
$dP/dV$ reaches its maximum value.
This finding indicates that, also at finite temperatures,
the $\gamma$-$\alpha$ volume-collapse transition is induced by
the rapid reduction of the Kondo temperature, which
occurs because the spin-orbit coupling effectively reduces the
$f$-level degeneracy from $14$ to $6$ at $V\simeq V_{\text{c}}$.

Note that,
while at zero temperature we find that $Z>0$ even at large volumes,
at $T>0$ we find that $Z_{5/2}=0$ for all
volumes larger than a critical value $V_{\text{M}}>V_{\text{c}}$,
which approaches $V_{\text{c}}$ when the temperature is increased.
This observation is suggestive respect to the considerations
of Ref.~\cite{Kondo-Hubbard-Held},
which partially reconciliate the HM and the KVC pictures
at finite temperatures.
On the other hand, we
believe that the second-order selective Mott transition
found in our calculations
does not reflect an actual physical effect,
but is an artifact of the SB approximation scheme.
More precisely, we do not expect that in cerium
a transition should be observed
as a function of the volume, but only as a function of the
pressure.
In any case, we point out that 
the thermodynamical properties or cerium are
determined exclusively by the free energy, which
does not display any singular behavior at $V_{\text{M}}$
in our calculations.

\begin{figure}
\begin{center}
\includegraphics[width=8.6cm]{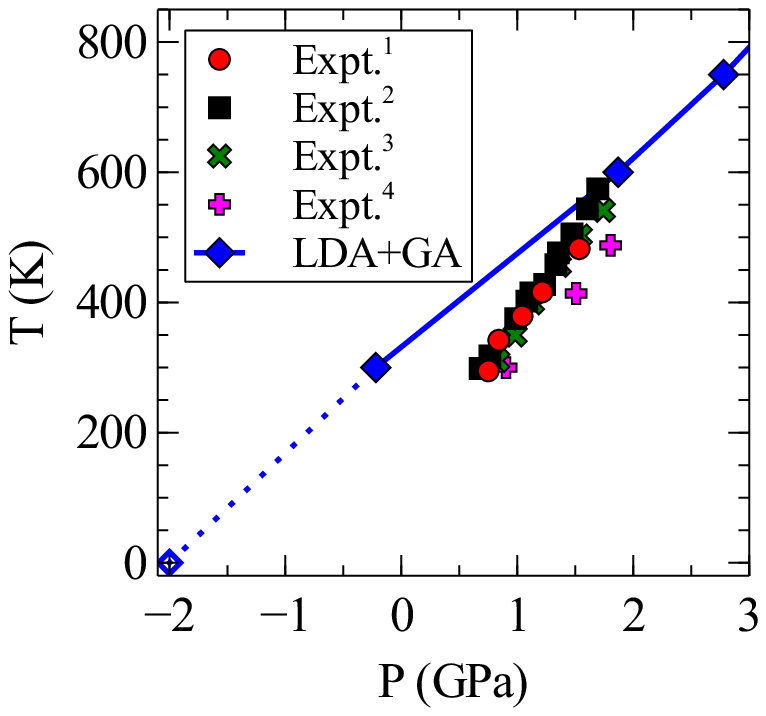}
\caption{Theoretical pressure-temperature
  phase diagram in comparison with the experimental data
  of Refs.~\cite{Lipp08,SwensonCe,Jayaraman,exptCe4}.
  The empty square at $T=0$ and negative pressures indicates 
  the signature of the transition identified by a maximum
  (but negative) value of $dP/dV$,
  while the other points indicate actual thermodynamical instabilities
  identified by a positive value of $dP/dV$.
}
\label{figure2}
\end{center}
\end{figure}
In Fig.~\ref{figure2} the 
theoretical pressure-temperature
phase diagram extrapolated from the free-energy curves
of Fig.~\ref{figure1} is shown in comparison with the
experimental data. 
Note that at zero temperature no
transition is observed in our calculations with $U=6\,eV$,
but only a signature identified by a maximum (but negative)
value of $dP/dV$, see Ref.~\cite{Our-Ce}.
The empty square at $T=0$ and negative pressures indicates 
this signature.
Remarkably, 
the behavior of the transition line in the pressure-temperature
phase diagram is in good agreement with the experiments, see Fig.~\ref{figure2}.
In particular, note that the critical pressure $P_{\text{c}}$
increases as a function of the temperature, and it is
positive for $T\gtrsim 300\,K$.
On the other hand, our calculations
do not reproduce the experimental behavior of the transition line
at high temperatures.
In fact, while experimentally the transition line ends at $T_{\text{c}}\simeq 500\,K$,
our calculations indicate that the transition line continues far beyond the
range of temperatures displayed in Fig.~\ref{figure2}, as 
a volume-collapse transition is observed
even at $T=1500\,K$.

\begin{figure}
\begin{center}
\includegraphics[width=8.6cm]{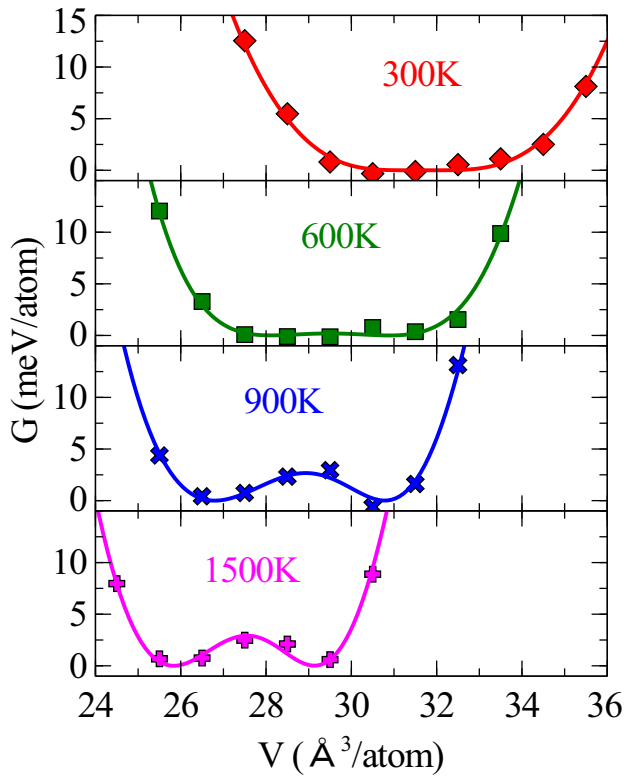}
\caption{Thermodynamical potentials
  $G=E-TS+P_{\text{c}}V$ as a function of the volume
  evaluated at different temperatures $T$ by fixing the corresponding
  critical pressures $P_{\text{c}}(T)$ of Fig.~\ref{figure2}.
  The continue curves are obtained by fitting the free energy
  data of Fig.~\ref{figure1} over the entire range of volumes
  using a one-dimensional smoothing spline fit with order $4$
  and smoothing factor $5\times 10^{-2}\,meV$,
  as implemented in the python class
  ``scipy.interpolate.UnivariateSpline''.
}
\label{figure3}
\end{center}
\end{figure}
In Fig.~\ref{figure3} are shown the thermodynamical potentials
$G=E-TS+P_{\text{c}}V$ as a function of the volume,
which are evaluated at different $T$ by fixing the corresponding
critical pressures $P_{\text{c}}(T)$.
Consistently with the definition of the common-tangent construction,
$G$ displays a double-minima structure,
indicating that the system is unstable at $P=P_{\text{c}}$.
Interestingly, the energy-barrier $\Delta$ between the $\alpha$ and $\gamma$ phases
is smaller than $4\,meV$ for all of the temperatures considered,
and it increases monotonically with $T$. 
Note also that $\Delta$ increases rapidly only at low temperatures,
while it remains almost unchanged between $T=900\,K$ and $1500\,K$.

We point out that, within the Born-Hoppenheimer approximation,
the function $G=E-TS+PV$ represents the effective
potential experienced by the cerium atoms at fixed temperature $T$
and pressure $P$.
This observation suggests that
the lattice vibrations --- which were not taken into account in
our calculations --- might be the actual
reason why the transition line ends at the finite temperature
$T_{\text{c}}\simeq 500\,K$.
In fact,  since the energy-barrier $\Delta$
is very small even at high temperatures,
it might be possible that, because of the 
temperature-induced lattice vibrations,
the atom-configurations
``sample'' simultaneously both the $\alpha$ and the $\gamma$ phase
not only at the critical pressure $P_{\text{c}}$ defined above,
but also within a small interval of pressures $P\simeq P_{\text{c}}$,
thus transforming the volume-collapse transition in a crossover
when the temperature is sufficiently high.

In conclusion, we have calculated from first principles
the phase diagram of fcc cerium, finding good agreement
with the experiments.
Our analysis suggests that the iso-structural pressure-induced
$\gamma$-$\alpha$ transition is induced by a rapid variation
of the Kondo temperature, which occurs because of
the interplay between the spin-orbit coupling
and the $f$-electron correlations.
While a signature of the volume-collapse transition appears
in the free energy already at low temperatures,
at higher temperatures this signature is enhanced because
of the entropic effects, and originates an actual thermodynamical
instability.
The behavior of the theoretical thermodynamical potential at constant
pressure indicates that the $\gamma$ and $\alpha$ phases 
are separated by a very small energy barrier at all temperatures.
This observation suggests
an appealing possible explanation of why the pressure-temperature
transition line ends at $T_{\text{c}}\simeq 500\,K$:
at $T>T_{\text{c}}$
the atom configurations might be able to overcome the above-mentioned
barrier and sample simultaneously the $\gamma$ and $\alpha$ phases
even slightly before the critical pressure,
thus transforming the $\gamma$-$\alpha$ transition in a crossover.
In order to clarify this point,
and thus answer the third of the questions listed in the
introduction of this work,
it would be interesting to perform molecular dynamics
simulations~\cite{parrinello_MD_crystals,MD_Pu-ZPYin}, e.g.,
using effective atomistic potentials extrapolated
from the free-energy curves of this work.
Alternatively, this issue could be investigated by performing 
direct Monte Carlo simulations (iterating LDA+SB calculations
of cerium with a large unit cell).

\section{ACKNOWLEDGMENTS}

We thank Kristjan Haule, Michele Fabrizio and XiaoYu Deng for useful discussions.
~N.L. and G.K. were supported by U.S. DOE Office of
Basic Energy Sciences under Grant No. DE-FG02-99ER45761.
~The collaboration was supported by the U.S. Department of Energy
through the Computational Materials and Chemical Sciences Network CMSCN.
Research at Ames Laboratory supported by the U.S.
Department of Energy, Office of Basic Energy Sciences,
Division of Materials Sciences and Engineering. 
Ames Laboratory is operated for the U.S. Department of Energy
by Iowa State University under Contract No. DE-AC02-07CH11358.

\bibliographystyle{apsrev}


\end{document}